# Nature of novel moiré exciton states in WSe$_2$/WS$_2$ heterobilayers

*Mit H. Naik[1, 2]\*, Emma C. Regan[1,2,3]\*, Zuocheng Zhang[1]\*, Yang-hao Chan[1,2,4], Zhenglu Li[1,2], Danqing Wang[1,3], Yoseob Yoon[1,2], Chin Shen Ong[1,2], Wenyu Zhao[1], Sihan Zhao[5], M. Iqbal Bakti Utama[1,2,6], Beini Gao[1], Xin Wei[1], Mohammed Sayyad[7], Kentaro Yumigeta[6], Kenji Watanabe[8,9], Takashi Taniguchi[8,9], Sefaattin Tongay[7], Felipe H. da Jornada[10], Feng Wang[1,2,11], Steven G. Louie[1,2]†*

[1]Department of Physics, University of California at Berkeley, Berkeley, California 94720, United States

[2]Materials Sciences Division, Lawrence Berkeley National Laboratory, Berkeley, California 94720, United States

[3]Graduate Group in Applied Science and Technology, University of California at Berkeley, Berkeley, California 94720, United States

[4]Institute of Atomic and Molecular Sciences, Academia Sinica, and Physics Division, National Center for Theoretical Sciences, Taipei 10617, Taiwan

[5]Interdisciplinary Center for Quantum Information, Zhejiang Province Key Laboratory of Quantum Technology and Device, State Key Laboratory of Silicon Materials, and Department of Physics, Zhejiang University, Hangzhou, 310027, China

[6]Department of Materials Science and Engineering, University of California at Berkeley, Berkeley, California 94720, United States

[7]School for Engineering of Matter, Transport and Energy, Arizona State University, Tempe, Arizona 85287, United States

[8]Research Center for Functional Materials, National Institute for Materials Science, 1-1 Namiki, Tsukuba 305-0044, Japan

[9]International Center for Materials Nanoarchitectonics, National Institute for Materials Science, 1-1 Namiki, Tsukuba 305-0044, Japan

[10]Department of Materials Science and Engineering, Stanford University, Stanford, California 94305, United States.

[11]Kavli Energy NanoSciences Institute at University of California Berkeley and Lawrence

Berkeley National Laboratory, Berkeley, California 94720, United States

* These authors contributed equally to this work

† Correspondence to: sglouie@berkeley.edu

**Abstract:**

Moiré patterns of transition metal dichalcogenide (TMD) heterobilayers have proven to be an ideal platform to host unusual correlated electronic phases, emerging magnetism, and correlated exciton physics. While the existence of novel moiré excitonic states is established[1–4] through optical measurements, the microscopic nature of these states is still poorly understood, often relying on empirically fit models. Here, combining large-scale first-principles *GW*-BSE calculations and micro-reflection spectroscopy, we identify the nature of the exciton resonances in $WSe_2/WS_2$ moiré superlattices, discovering a surprisingly rich set of moiré excitons that cannot be even *qualitatively* captured by prevailing continuum models. Our calculations reveal moiré excitons with distinct characters, including modulated Wannier excitons and previously unindentified intralayer charge-transfer excitons. Signatures of these distinct excitonic characters are confirmed experimentally via the unique carrier-density and magnetic-field dependences of different moiré exciton resonances. Our study highlights the highly non-trivial exciton states that can emerge in TMD moiré superlattices, and suggests novel ways of tuning many-body physics in moiré systems by engineering excited-states with specific spatial characters.

Moiré superlattices of two-dimensional (2D) materials have emerged as a new platform to explore intriguing electronic phenomena including unconventional superconductivity, ferromagnetism, and tunable Wigner crystal and Mott insulating states[5–10]. Excitonic physics in TMD moiré superlattices have also garnered great interest, where multiple new resonances in the photoluminescence and absorption spectra can emerge in a moiré superlattice [1–4,11–13] and novel moiré exciton based photonic devices has been demonstrated[14]. For example, in the $WSe_2/WS_2$ heterostructure, the $WSe_2$ A exciton peak splits into multiple emergent peaks[1,10,15] when the twist-angle between the two layers is reduced to less than ~2°. This splitting is directly correlated to the formation of a large-area moiré superlattice at these small twist angles.

While the presence of moiré excitons is well established experimentally, the microscopic nature of these moiré excitons remains a mystery because, so far, optical probes do not have the spatial resolution to resolve the electron and hole positions of individual exciton states in the superlattice unit cell. To date the measured moiré exciton features have only been explained using a continuum model[16,17], which hypothesizes an effective moiré potential modulating the center-of-mass motion of the pristine exciton. This model assumes that the intrinsic nature of the exciton does not change in the moiré superlattice—i.e., the internal spatial electron-hole correlations are unchanged from those of pristine single layers. This assumption is likely to fail in real moiré systems, where the electrons and holes can experience different moiré potential and exhibit very different spatially modulated moiré flat bands[18–25]. We show indeed this is the case in $WSe_2/WS_2$ heterobilayers, leading to the discovery of moiré exciton of novel characteristics.

Parameter-free *GW* plus Bethe-Salpeter equation[26,27] (*GW*-BSE) calculations, which provide unparalleled insight into the atomistic nature of excitons in the 2D TMDs[28], have not been attempted so far for large-area moiré systems due to the computational challenges associated with a large number of atoms in the moiré unit cell. First-principles calculations based on time-dependent density functional theory with conventional exchange-correlation potentials[29] do not capture the complicated dielectric screening in quasi-2D semiconductors[30] which is important to correctly describe exciton states in 2D materials[28,30]. Here we develop a novel scheme to solve the Bethe-Salpeter equation of the entire moiré superlattice, which enables us to study the microscopic electron-hole distributions of different moiré exciton states. Combining *ab initio* predictions with micro-reflection spectroscopy, we establish the nature of various moiré exciton resonances in rotationally aligned $WSe_2/WS_2$ heterostructures. Our calculation reveals that different moiré excitons are formed by the moiré superlattice and exhibit distinct characters. Notably, the lowest energy moiré resonance (peak I)[1] is described by a Wannier-type exciton with the electron tightly correlated to the hole, while the highest energy moiré resonance (peak III) is characterized by an intralayer charge-transfer exciton with the electron and hole densities spatially separated by ~5 nm in the moiré superlattice with a binding energy that is nearly 200 meV. To our knowledge, this is the first identification of such a strong intralayer charge-transfer exciton in 2D materials. The predicted different characters of the moiré excitons are confirmed experimentally through their dependences on carrier doping and magnetic field in the moiré heterostructure.

The zero-twist-angle $WSe_2/WS_2$ heterostructure forms a moiré superlattice with an 8 nm lattice constant due to a slight mismatch (~4%) in the atomic lattice constants of $WS_2$ and $WSe_2$ (Fig. 1a). A single moiré unit cell contains 3903 atoms. Driven by the energetics of the various stacking configurations in the moiré pattern[18,24,31], the superlattice undergoes a structural reconstruction which leads to a 3D buckling and a redistribution of strains in each layer[19] (Fig. 1b). To capture quantitatively from first-principles the effective potential used in the continuum model[16], we compute the variation in WSe$_2$ band gap as a result of the local strain in the reconstructed layer. This analysis is used to obtain the effective potential in the continuum model for the exciton center of mass with a variation of ~90 meV (Fig. 1c).

Previous studies have established flat moiré bands for individual electrons and holes in WSe$_2$/WS$_2$ moiré bilayers[19,32]. The wavefunctions of the two layers do not hybridize significantly around the $K$ point in the Brillouin zone (BZ) due to their large energy separation and small spatial extent in the out-of-plane direction[19,33–35] (being dominated by the W $d$ orbital character). Since the WSe$_2$ layer has a smaller band gap, the low-energy reflection contrast spectrum is dominated by the WSe$_2$ intralayer excitons, which are well separated from the higher-energy WS$_2$ intralayer excitons[1].

Figure 1d compares the $GW$ electronic band structure of pristine and the strained WSe$_2$ superlattice in the moiré BZ. The strained WSe$_2$ has flatter bands and the moiré lattice potential opens gaps at the moiré BZ edges. The inhomogeneous strains lead to a spatial modulation of the electronic wavefunctions at the valence and conduction band edges (Fig. 1e,f). Remarkably, the WSe$_2$ band states around the $K$ point are dominated by the strain redistribution in the WSe$_2$ layer as opposed to interlayer coupling[19] (see supplementary information, Fig. S2). As a result, we can obtain the intralayer moiré exciton wavefunctions and energies by solving the BSE in a strained WSe$_2$ superlattice, which makes the problem more tractable theoretically.

We first write the exciton wavefunction as a sum over valence to conduction transitions of multiple bands in the moiré BZ:

$$\chi_S(\mathbf{r}_e, \mathbf{r}_h) = \sum_{vc\mathbf{k}_m} A^S_{vc\mathbf{k}_m} \psi_{c\mathbf{k}_m}(\mathbf{r}_e) \psi^*_{v\mathbf{k}_m}(\mathbf{r}_h) \qquad (1)$$

where $\psi_{ck_m}(r_e)$ and $\psi_{vk_m}(r_e)$ are the conduction ($c$) and valence ($v$) single-particle wavefunctions at the $k_m$ point in the moiré BZ, respectively. $S$ is the exciton quantum number, $r_e$ and $r_h$ are the electron and hole coordinates. The BSE is an effective two-particle equation that includes an interaction kernel[26], responsible for the attractive and exchange-like repulsive interaction between the electron and hole, which leads to the formation of excitons in the system. The computation of the kernel, which involves an integral over the electron and hole coordinate of the band states in the superlattice, is challenging due to the large area of the moiré superlattice. To alleviate this computational burden, we develop a Pristine Unit-cell Matrix Projection (PUMP) method, which is based on expanding the moiré cell band-state wavefunctions in terms of the pristine unit cell wavefunctions, as inspired by the band unfolding technique[36–38]. We first expand the valence and conduction wavefunctions of the strained WSe$_2$ ($|\psi\rangle$) in a basis of valence and conduction states of pristine WSe$_2$ ($|\Phi\rangle$) in the moiré BZ, respectively: $|\psi_{vk_m}\rangle = \sum_i a_i^{vk_m} |\Phi_{ik_m}^{val}\rangle$; $|\psi_{ck_m}\rangle = \sum_i a_i^{ck_m} |\Phi_{ik_m}^{cond}\rangle$, where $|\Phi_{ik_m}\rangle$ correspond to the pristine WSe$_2$ wavefunctions folded from the unit-cell BZ to the $k_m$ point in the moiré BZ. The projection ($\sum_i |a_i^{vk_m}|^2$ or $\sum_i |a_i^{ck_m}|^2$) of the moiré states is greater than 95%, indicating that pristine WSe$_2$ is a good basis for the expansion (see supplementary information Fig. S4). The PUMP expansion above is a coherent superposition of various Bloch states corresponding to different k-points in the unit-cell BZ. The complicated modulation of the strained WSe$_2$ wavefunction in the superlattice is completely captured with this treatment. Using this expansion, the strained superlattice electron-hole kernel matrix elements can be accurately approximated in terms of the pristine WSe$_2$ kernel matrix elements:

$$\langle \psi_{ck_m}\psi_{vk_m}|K^s|\psi_{c'k'_m}\psi_{v'k'_m}\rangle$$

$$= \sum_{i,j,p,q} a_i^{ck_m*} a_j^{vk_m*} a_p^{c'k'_m} a_q^{v'k'_m} \left\langle \phi_{\alpha k_{uc}^1}^{cond} \phi_{\beta k_{uc}^2}^{val} \Big| K^p \Big| \phi_{\gamma k_{uc}'^1}^{cond} \phi_{\eta k_{uc}'^2}^{val} \right\rangle \delta_{k_{uc}'^2 - k_{uc}'^1, k_{uc}^2 - k_{uc}^1} \quad (2)$$

where $K^s$ and $K^p$ are the kernel corresponding to strained WSe$_2$ and pristine WSe$_2$, respectively. The kernel of the strained superlattice is thus expressed in terms of a coherent superposition of unit-cell kernel matrix elements. This expansion makes the construction of the BSE kernel computationally faster by 6 orders of magnitude.

Our reflection contrast measurement of the hBN-encapsulated rotationally aligned WSe$_2$/WS$_2$ heterostructure (Fig. 2a, supplementary information) shows three dominant excitonic peaks (I, II and III), as opposed to a single A exciton peak in monolayer WSe$_2$[1]. We compare this observed spectrum to the absorbance spectra calculated using our *ab initio GW*-BSE method with PUMP expansion and the continuum model approach (Fig. 2b and c). The *ab initio* computed excitonic peaks and their relative oscillator strengths are in excellent agreement with the measurements. On the other hand, the continuum model based on the *ab initio* effective potential predicts only two dominant excitonic peaks (I$_m$ and II$_m$). We note the resonance energies do not match up exactly between the *ab initio* GW-BSE spectrum and experimental spectrum, which could be due to the use of an isolated monolayer in our calculations, as opposed to an encapsulated heterostructure in the measurement (environmental screening is known to give rise to redshift in the exciton energies especially for the higher energy states), or due to external strain in the experimental sample.

To better understand the nature of the moiré excitons, we plot the electron charge density of the exciton (Eq. (1)) for a fixed hole coordinate, $\boldsymbol{r}_h$, in the superlattice. In Fig. 2d and 2e, $\boldsymbol{r}_h$ is placed at three specific W atom positions along the (11) direction, corresponding to the three high-symmetry stacking sites in a bilayer, labelled AA, $B^{Se/W}$, and $B^{W/S}$. For the case of the A exciton of a pristine WSe$_2$ layer, the electron density always centers and peaks around the hole, forming a Wannier exciton, with no preference for specific sites as the hole goes from one pristine unit cell to another. On the other hand, the modulation of the band states (Fig. 1e,f) in the strained superlattice leads to a dramatic change in the moiré exciton states. For exciton peak I from *ab initio* GW-BSE, the electron density is maximum at the AA stacking site when the hole position is also at the same stacking site (Fig. 2d). This exciton thus possesses a modulated Wannier character with the electron and hole strongly correlated in their positions. The peak III exciton, in contrast, has maximum electron density at the nearby AA stacking sites when the hole coordinates are at the $B^{Se/W}$ site region. Moreover, when the hole is at the AA stacking site, the electron density is negligible (Fig. 2d). This is a clear indication that for this state the electron and hole densities are separated in the WSe$_2$ layer, forming a charge-transfer exciton with electron-hole separation of about 5 nm. Charge-transfer excitons are typically formed in molecular solids with acceptor/donor molecules or across an interface due to a type II band alignment between two materials. The exciton we observe here, on the other hand, is within a single layer. The peak II exciton has a mixed character (see supplementary information Fig. S8). However, the continuum model is intrinsically unable to predict a moiré charge-transfer exciton here since the electron and hole degrees of freedom are not explicitly included (Fig. 2e), so the resulting excitons always retain the Wannier character of the pristine WSe$_2$ A exciton -- as the moiré excitons are constrained to be expressed as a 1*s* Wannier exciton that is modulated through its center-of-mass coordinate $R = (m_e r_e + m_h r_h)/(m_e + m_h)$ in space (here $m_e$ and $m_h$ are the electron and hole effective masses, respectively). Our findings demonstrate the need to go beyond the continuum model to even *qualitatively* describe the nature of excitons in moiré superlattices.

The distinct nature of these moiré excitons can be experimentally probed through their carrier doping dependence, with and without a magnetic field, in a WSe$_2$/WS$_2$ moiré heterostructure (see supplementary information for device information and experimental methods). We will focus on the moiré peaks I and III, which have well-defined Wannier and charge-transfer characteristics, respectively. Figure 3a shows the evolution of the WSe$_2$ intralayer moiré exciton resonances as a function of electron and hole doping in a near-zero-twist-angle WSe$_2$/WS$_2$ heterostructure[1,7,15], which reveals distinctly different behavior compared with the exciton transitions in a large-twist-angle heterostructure (Fig. 3b). In both twist-angle cases, the doping electrons reside in the WS$_2$ layer and the doping holes reside in the WSe$_2$ layer due to the type II band alignment.

The doping-dependent absorption features in the large-twist-angle WSe$_2$/WS$_2$ heterostructure can be understood straightforwardly[39]: The hole doping in the WSe$_2$ layer leads to the formation of the positively charged trion resonance and suppression of the the 1s exciton transition in the WSe$_2$ layer, while the electron doping in the WS$_2$ layer leads to a slight red-shift in the peak due to screening from the free carriers[40,41]. The doping-dependence of the moiré exciton absorption spectra in aligned WSe$_2$/WS$_2$ heterobilayer, however, exhibit several puzzling behaviors that are at first sight highly counter-intuitive. First, the dominating moiré exciton peak I in the WSe$_2$ layer shows negligible energy shift when holes are doped into the same WSe$_2$ layer with no apparent trion formation, while peak III is strongly affected. Second, moiré peaks I and

III in the WSe$_2$ layer both exhibit strong apparent changes when electrons are doped into the WS$_2$ layer. These large shifts are difficult to be explained solely by considering screening effects from the doping carriers. We show here that this puzzling doping dependence is a manifestation of the unusual moiré exciton states, which couple differently to the modulated doping electrons and holes in the moiré superlattice.

Upon hole doping of the moiré superlattice, the doping holes have maximum density at the B$^{Se/W}$ site of the WSe$_2$ layer (Fig. 1f). However, both the electron and hole of moiré exciton I are at the AA stacking site (Fig. 2d and 3c). As a result, the strongest WSe$_2$ moiré exciton I is minimally affected by hole-doping, even though the holes are in the same WSe$_2$ layer. Moiré peak III, on the other hand, is strongly modified by hole doping because the constituent hole density of the exciton is maximum at the B$^{Se/W}$ site (Fig. 2d and 3c), the same position as the doping holes. Here we distinguish two scenarios when the doping holes are present only in a fraction of the moiré lattice sites. In scenario A (in Fig. 3d), the doping hole (the red dot) and the photo-excited constituent hole (the orange dot) occupy the same B$^{Se/W}$ site. Coulomb repulsion shifts the exciton transition to much higher energy and the peak broadens. Pauli blocking of the holes also makes the transition weaker. The broadening and suppression makes the resonance in scenario A not observable experimentally. In scenario B (in Fig. 3d), the doped hole and constituent hole occupy adjacent B$^{Se/W}$ sites. The Coulomb interaction between the doping hole and the moiré exciton dipole will decrease the total energy and lead to a redshift in peak III (dashed line in Fig. 3a). Combination of scenario A and B accounts for the observed overall redshift and suppression of peak III upon hole doping.

Upon electron doping, the doping electrons have the maximum density at the B$^{Se/W}$ site in the WS$_2$ layer[32], and are not on the WSe$_2$ layer. If electron doping in the WS$_2$ layer only has minimum effect on the tightly bound 1*s* exciton in the WSe$_2$ layer in a large twist angle heterostructure (Fig. 3b), how can it exert strong effect on intralayer moiré excitons in an aligned sample? The doping electrons at the WS$_2$ B$^{Se/W}$ site should have negligible effect on the moiré peak I, where the constituent electrons and holes occupy the AA site in WSe$_2$ (Fig. 3e). For the moiré peak III, the constituent hole (electron) is at the B$^{Se/W}$ (AA) site in WSe$_2$ (Fig. 3e). The extremely large dipole of charge-transfer exciton in moiré peak III allows for strong Coulomb coupling to the doping electrons in the WS$_2$ layer. Again, we distinguish two scenarios when doped electrons are present only in a fraction of the moiré lattice sites. In scenario A (in Fig. 3e), the doped electron (blue dot) and the constituent hole (orange dot) occupy the same B$^{Se/W}$ site. The attractive electron-hole interaction leads to a dramatic reduction of the exciton energy and the emergence of a new peak very close to moiré peak I, which can partially hybridize with the moiré peak I through near-resonant coupling. We can estimate the redshift by computing the electrostatic interaction of the doped electron density in the WS$_2$ layer with the charge-transfer exciton in the WSe$_2$ layer. Using the computed charge densities of the electrons and holes in the moire heterostructure, we estimate a redshift of ~120 meV (see supplementary information for details), in good agreement with the experimental observation. In scenario B (in Fig. 3e), the doping electron and constituent hole occupy adjacent B$^{Se/W}$ sites. The Coulomb interaction between the doping electron and the moiré exciton dipole increases the total energy and leads to a blueshift in peak III, as observed experimentally (dotted line in Fig. 3a).

We can futher confirm the distinct nature of the moiré exciton states through their response to a perpendicular magnetic field. Fig. 4 displays the reflection contrast spectra of WSe$_2$/WS$_2$ moiré superlattice (Fig. 4a) and a WSe$_2$ monolayer (Fig. 4b) for both left and right circular light under a perpendicular magnetic field at $B = 6$ T. Here we use right circular light $\sigma+$ to probe excitons in the K valley and left circular light $\sigma-$ to probe excitons in the K' valley[42–44].

At charge neutral, the monolayer $WSe_2$ A exciton (Fig. 4b) and the three moiré excitons in the aligned heterostructure (Fig. 4a) shows a weak Zeeman splitting (i.e., the difference in peak positions) of the exciton transitions. Upon hole doing, each exciton feature exhibits a distinct evolution depending on its interaction with the doping holes. At the magnetic field of 6 T, the doping holes are completely valley polarized at the K' valley[45]. Therefore, the absorption differences for K valley ($\sigma+$) and K' valley ($\sigma-$) excitons at finite hole doping can be understood based on the exchange interaction between the K'-valley doping holes and the constituent holes in different excitons.

The exchange interaction is strong in monolayer $WSe_2$, which leads to very different spectra for K and K' valley excitons upon hole doping (Fig. 4b). The K'-valley doping holes and K valley exciton can form bound trion state at a lower energy[46,47]. Such trion state is not present for K'-valley exciton due to the exchange interaction between the spin- and valley-polarized holes.

In contrast, the moiré excitons in the aligned heterostructure exhibit very different behaviors. Moiré peak I shows much smaller difference between the K and K' valley excitons upon hole doping at K' valley compared with that in monolayer $WSe_2$, indicating a very weak exchange interaction between the exciton I and the doping holes. This behavior is a manifestation of the hole and exciton separation: the doping holes have largest charge density at the $B^{Se/W}$ site of the $WSe_2$ layer (Fig. 1f), while both the electron and hole of moiré exciton I are at the AA stacking site (Fig. 3c). Moiré peak II shows relatively strong difference in the K and

K' exciton absorption with doping holes at K' valley. This observation is consistent with partial spatial overlap of the constituent hole in moiré exciton II and the doping hole at the $B^{Se/W}$ site, which results in significant exchange interaction. Moiré peak III shows a strongly diminished peak upon hole doping with no valley polarization dependence (beyond the Zeeman effect). The residue moiré peak III comes from scenario B in Fig. 3d, where the doped hole and constituent hole of exciton III occupy adjacent $B^{Se/W}$ sites. Because the doping holes have no spatial overlap with the exciton, the exchange interaction is negligible.

Our study shows that novel exciton states with highly unusual electron-hole correlations, including intralayer charge-transfer excitons, can be present in TMD moiré superlattices. They can give rise to dramatically enhanced electron-exciton interactions, as demonstrated by the strong coupling of an electron in one layer with a charge-transfer exciton in the other layer. The microscopic understanding of the moiré exciton wavefunction, as demonstrated in our work, could lead to new ways for electrical control of moiré excitons for nanophotonic devices.

**Acknowledgements**: This work was primarily supported by the Director, Office of Science, Office of Basic Energy Sciences, Materials Sciences and Engineering Division of the US Department of Energy under the van der Waals heterostructure program (KCWF16), contract number DE-AC02-05CH11231 which provided the theoretical PUMP formulation and experimental measurements, supported by the Center for Computational Study of Excited-State Phenomena in Energy Materials (C2SEPEM) at LBNL, funded by the U.S. Department of Energy, Office of Science, Basic Energy Sciences, Materials Sciences and Engineering Division under Contract No. DE-AC02-05CH11231, as part of the Computational Materials Sciences Program which provided advanced codes and simulations, and supported by the Theory of Materials Program (KC2301) funded by the U.S. Department of Energy, Office of Science, Basic Energy Sciences, Materials Sciences and Engineering Division under Contract No. DE-AC02-05CH11231 which provided conceptual and symmetry analyses of excitons. Computational resources were provided by National Energy Research Scientific Computing Center (NERSC), which is supported by the Office of Science of the US Department of Energy under contract no. DE-AC02-05CH11231, Stampede2 at the Texas Advanced Computing Center (TACC), The University of Texas at Austin through Extreme Science and Engineering Discovery Environment (XSEDE), which is supported by National Science Foundation under grant no. ACI-1053575 and Frontera at TACC, which is supported by the National Science Foundation under grant no. OAC-1818253.

**Author contributions:** S.G.L. conceived the project; M.H.N., Y.C., Z.L., C.S.O, F.H.J. and S.G.L. developed the theory and performed the GW-BSE calculations; E.C.R., Z.Z., D.W, Y.Y. carried out optical measurements and analysis. E.C.R., Z.Z., D.W, W.Z., S.Z., M.I.B.U., B.G., X.W. fabricated van der Waals heterostructures. M.S., K.Y., M.B., S.T. grew WSe2 and WS2 crystals. K.W. and T.T. grew hexagonal boron nitride crystals.

**Competing interests:** The authors declare no competing interests.

**Data and materials availability:** The data that support the findings of this study are available from the corresponding author upon reasonable request.

**Corresponding author:** Correspondence and requests for materials should be addressed to sglouie@berkeley.edu

Figures:

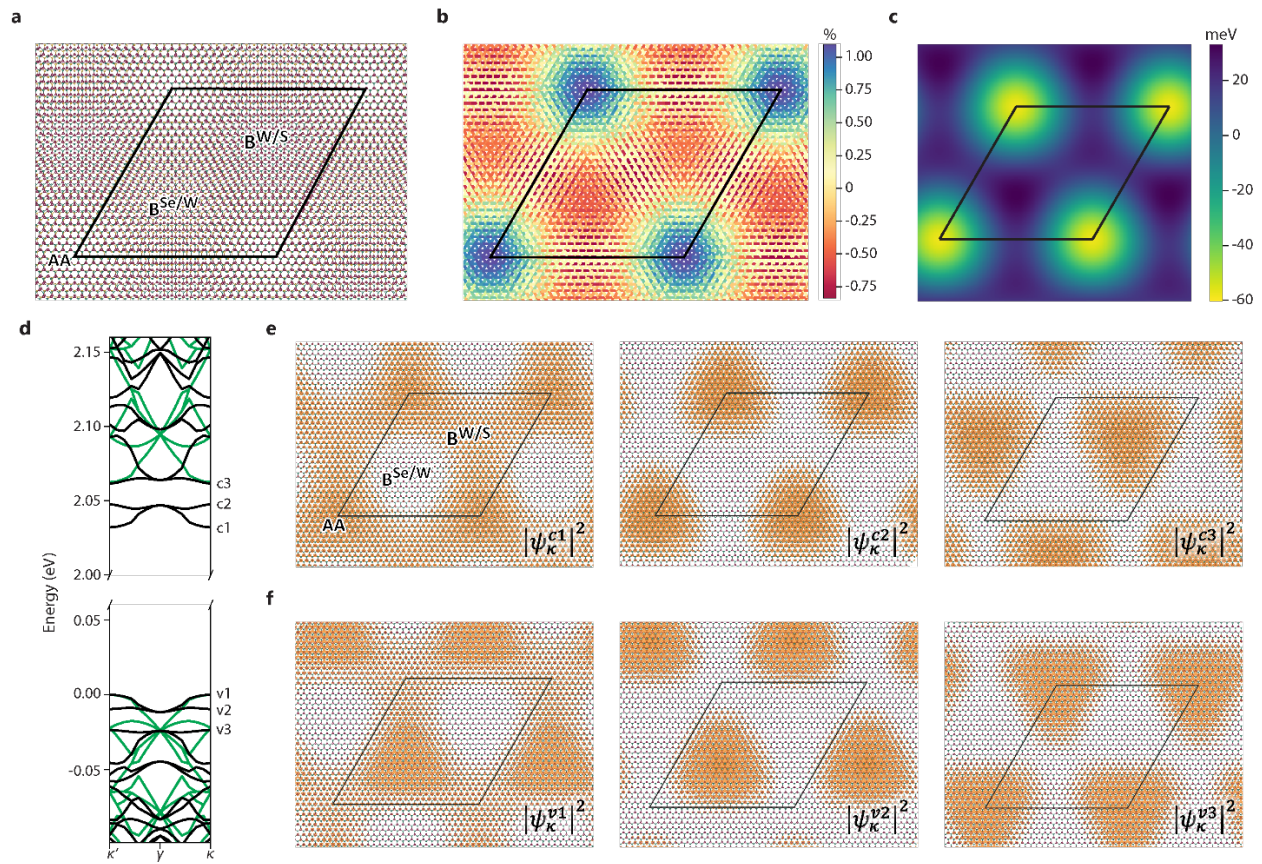

**Fig. 1 | Reconstruction of the moiré superlattice.** (**a**) Moiré superlattice formed in rotationally aligned WSe$_2$/WS$_2$ heterostructure with the superlattice cell outlined in black and the high-symmetry positions labelled. (**b**) Strain distribution in the WSe$_2$ layer as a result of structural reconstruction. The strain is plotted as a percentage change in the local lattice parameter from equilibrium. (**c**) The effective moiré potential used in the continuum model, computed as a variation in the WSe$_2$ band gap as a function of local strain in the superlattice. (**d**) GW band structure of pristine (green line) and strained (black line) WSe$_2$ plotted in the moiré BZ. The strain pattern (shown in **b**) leads to energy-separated flat bands at the valence and conduction band edges. (**e** and **f**) Electronic wavefunctions of states at the conduction (**e**) and valence (**f**) band edges are modulated due to the strain. The wavefunction squared of states at the κ point in the moiré BZ as labeled in **d** are plotted with an isosurface of 8% of the maximum.

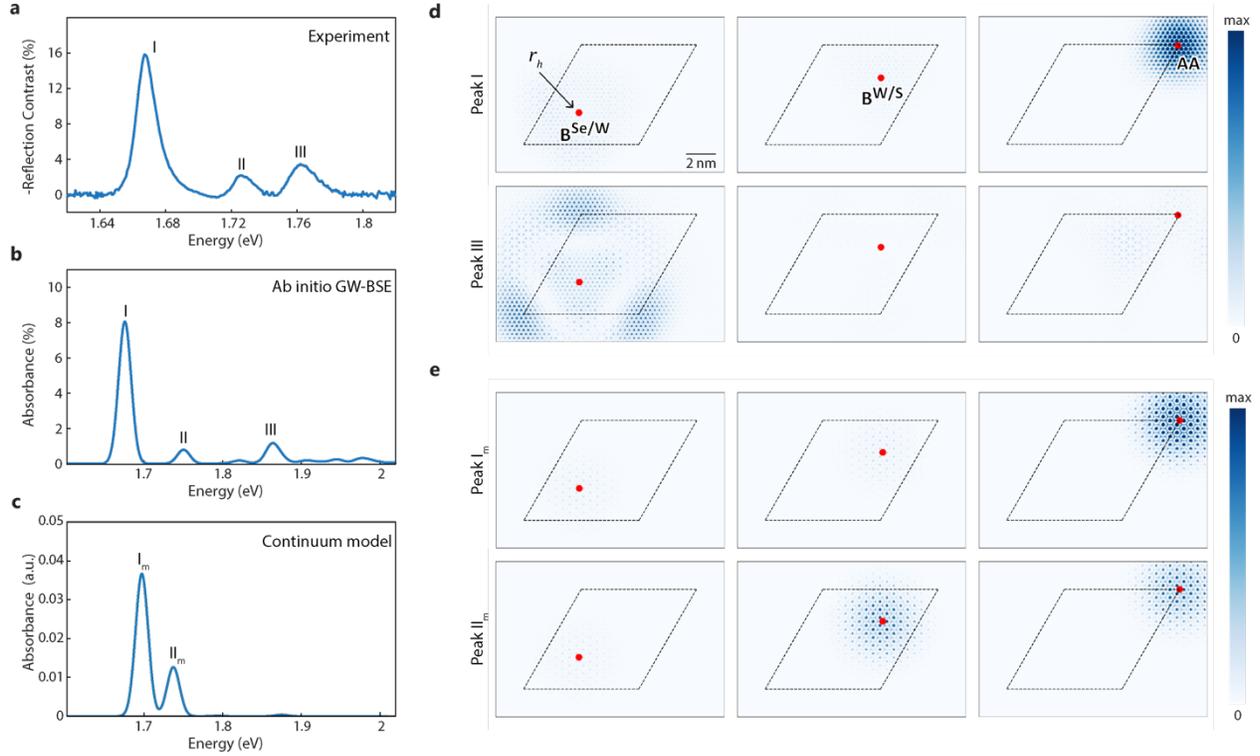

**Fig. 2 | Optical spectrum and nature of intralayer excitons in rotationally aligned** $WSe_2/WS_2$. (**a**) Experimental reflection contrast spectrum of a rotationally aligned $WSe_2/WS_2$ heterostructure. (**b**) *Ab initio GW*-BSE absorbance spectrum of the inhomogeneously strained $WSe_2$ moiré superlattice. (**c**) Absorbance spectrum obtained by solving the continuum model with an effective potential calculated from *ab initio* as described in Fig. 1c. (**d**) Distribution of the electron charge density of the *ab initio* computed exciton, $|\chi_S(r_e, r_h)|^2$, forming peak I and peak III, as a function of fixed hole position (red dot) in the moiré unit cell. The two excitons are distinctly different: a modulated Wannier-like exciton for peak I, and a charge-transfer exciton for peak III. (**e**) Similar plot for the peaks $I_m$ and $II_m$ from the continuum model. Here, the exciton center-of-mass amplitude for peak $I_m$ and $II_m$ obtained by solving the continuum model is used to modulate the pristine $WSe_2$ A exciton in the plots.

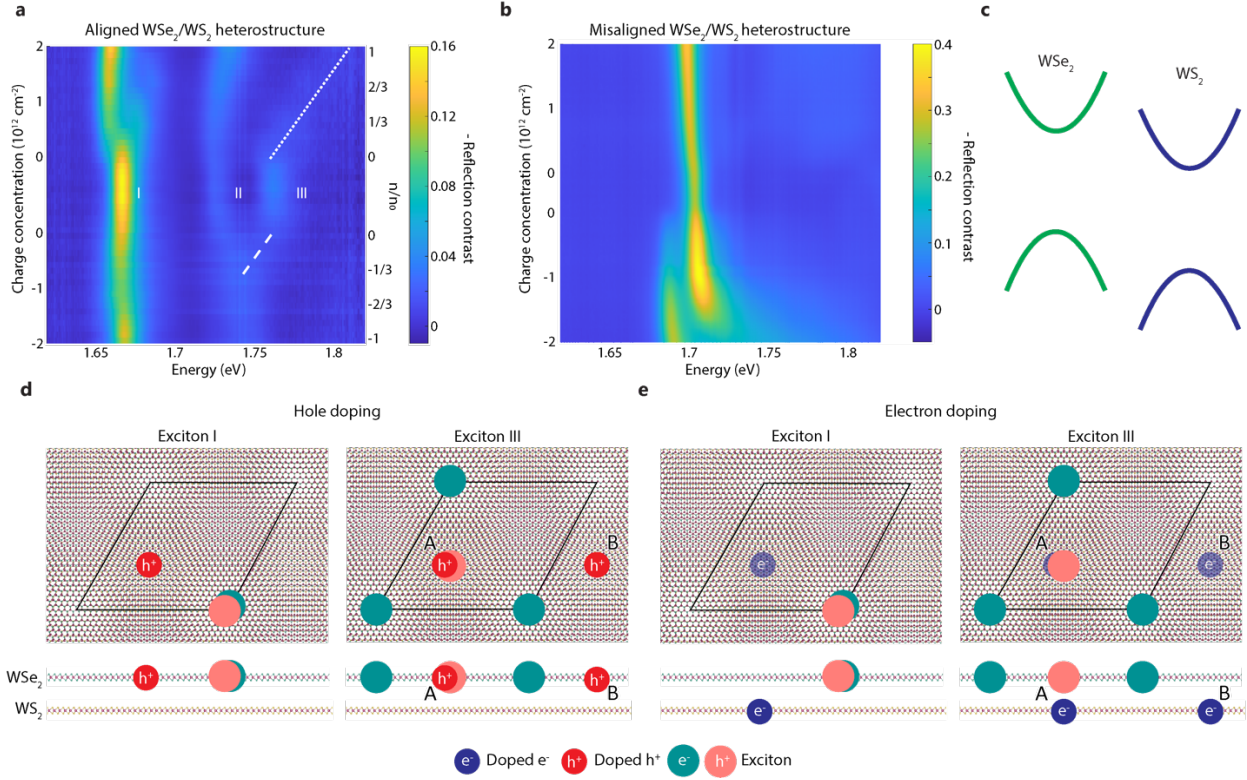

**Fig. 3 | Behavior of** $WSe_2/WS_2$ **moiré excitons under charge doping.** (**a**) Experimental doping-dependent reflection contrast spectrum of a $WSe_2/WS_2$ moiré superlattice shows distinct behavior for $WSe_2$ moiré excitons I, II, and III. Negative (positive) charge concentration refers to hole (electron) doping. The right vertical axis diplays the charge concentration per moiré unit cell ($n_0 = 1.8 \times 10^{12}$ cm$^{-2}$). The redshift (blueshift) of peak III under hole (electron) doping is highlighted with dashed (dotted) white lines. (**b**) Experimental doping-dependent reflection contrast spectrum of a reference misaligned $WSe_2/WS_2$ heterostructure. The $WSe_2$ exciton interacts strongly with doping holes in the $WSe_2$ layer, forming a trion state, but weakly with doping electrons in the $WS_2$ layer. (**c**) Schematic of band alignment. (**d** and **e**) Schematic of moiré exciton I (left) and III (right) under (**d**) hole and (**e**) electron doping from both top-down (top) and side (bottom) views. Exciton I is a tightly bound exciton at the AA position, while exciton III is a charge-transfer exciton with the hole at the $B^{Se/W}$ position and electron density at the surrounding AA positions. Doping electrons and holes both reside at the $B^{Se/W}$ positions. The relative position of the doping charge and the constituent exciton charge determine the unique doping dependence of each state. In the case of exciton III, we consider two scenarios: in scenario A, the doping charge is at the same $B^{Se/W}$ site as the constituent hole, and in scenario B, the doping charge is at the adjacent $B^{Se/W}$ site.

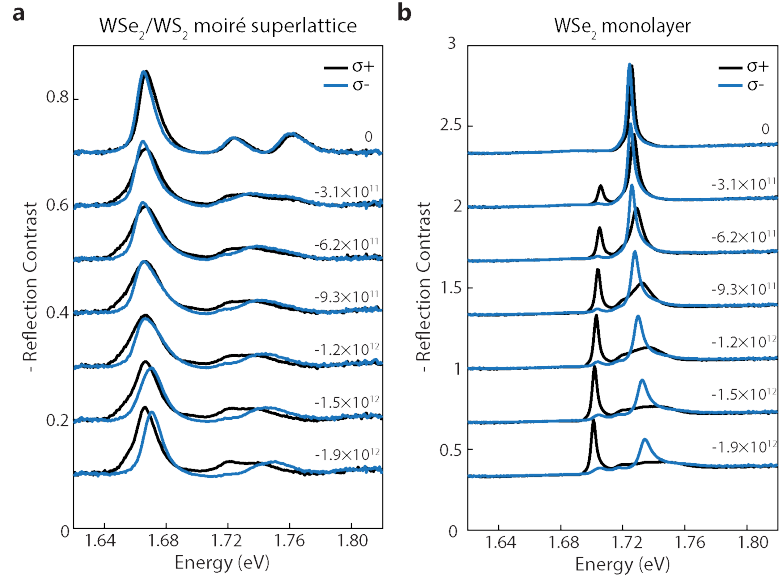

**Fig. 4 | Doping dependence of moiré excitons under magnetic field.** Hole doping dependent reflection contrast spectra of $WSe_2/WS_2$ moiré superlattice (**a**) and $WSe_2$ monolayer (**b**) for both left and right circularly polarized light under a perpendicular magnetic field at $B = 6$ T. Hole density is labeled for each set of spectra. (a) Magnetoreflection contrast of $WSe_2/WS_2$ moiré superlattice. Exciton I and III show relatively weak dependence on the exciton valley due to the spatial separation between doping holes at the K' valley and the constituent holes. Exciton II shows a relatively strong difference in behavior between K and K' valley excitons due to the partial overlap between doping holes at the K' valley and the exciton. (b) Magnetoreflection contrast of $WSe_2$ monolayer for comparison. The monolayer $WSe_2$ A exciton is strongly modified by the doping holes at the K' valley due to the large exchange interaction. K' valley doping holes form trion states with the K valley exciton, resulting in an absorption peak at lower energy.

Supplementary information for

# Nature of novel moiré exciton states in WSe$_2$/WS$_2$ heterobilayers

Mit H. Naik[1,2*], Emma C. Regan[1,2,3*], Zuocheng Zhang[1*], Yang-hao Chan[1,2,4], Zhenglu Li[1,2], Danqing Wang[1,3], Yoseob Yoon[1,2], Chin Shen Ong[1,2], Wenyu Zhao[1], Sihan Zhao[5], M. Iqbal Bakti Utama[1,2,6], Beini Gao[1], Xin Wei[1], Mohammed Sayyad[7], Kentaro Yumigeta[6], Kenji Watanabe[8,9], Takashi Taniguchi[8,9], Sefaattin Tongay[7], Felipe H. da Jornada[10], Feng Wang[1,2,11], Steven G. Louie[1,2]†

[1]Department of Physics, University of California at Berkeley, Berkeley, California 94720, United States

[2]Materials Sciences Division, Lawrence Berkeley National Laboratory, Berkeley, California 94720, United States

[3]Graduate Group in Applied Science and Technology, University of California at Berkeley,

Berkeley, California 94720, United States

[4]Institute of Atomic and Molecular Sciences, Academia Sinica, and Physics Division, National Center for Theoretical Sciences, Taipei 10617, Taiwan

[5]Interdisciplinary Center for Quantum Information, Zhejiang Province Key Laboratory of Quantum Technology and Device, State Key Laboratory of Silicon Materials, and Department of Physics, Zhejiang University, Hangzhou, 310027, China

[6]Department of Materials Science and Engineering, University of California at Berkeley, Berkeley, California 94720, United States

[7]School for Engineering of Matter, Transport and Energy, Arizona State University, Tempe,

Arizona 85287, United States

[8]Research Center for Functional Materials, National Institute for Materials Science, 1-1 Namiki,

Tsukuba 305-0044, Japan

[9]International Center for Materials Nanoarchitectonics, National Institute for Materials Science,

1-1 Namiki, Tsukuba 305-0044, Japan

[10]Department of Materials Science and Engineering, Stanford University, Stanford, California 94305, United States.

[11]Kavli Energy NanoSciences Institute at University of California Berkeley and Lawrence

Berkeley National Laboratory, Berkeley, California 94720, United States

\* These authors contributed equally to this work

† Correspondence to: sglouie@berkeley.edu

## S1. Details of DFT and GW-BSE studies

### S1.1 Moiré reconstruction and density functional theory calculations

The moiré superlattice for 0° twist is constructed using a 25 × 25 supercell of $WSe_2$ and 26 × 26 supercell of $WS_2$ to ensure a mismatch of less than 1% between the moiré lattice vectors of the two layers. The lattice constant of pristine $WSe_2$ and $WS_2$ used in the calculation is 3.32 Å and 3.19 Å, respectively. The superlattice is encapsulated by a layer of hBN on either side for the structural relaxation step to simulate the experimental configuration. Structural relaxation of the hBN/$WSe_2$/$WS_2$/hBN heterostructure is performed using forcefields as implemented in the LAMMPS[1] package with the help of the TWISTER[2] code. The intralayer forcefield used in the calculation is Stillinger-Weber[3] for the transition metal dichalcogenides (TMDs) and Tersoff[4] for hBN. Interlayer interactions are included using the Kolmogorov-Crespi[5,6] forcefield. The force tolerance for the relaxation is set to be $10^{-4}$ eV/Å. The ground-state electronic structure of the reconstructed $WSe_2$ layer is studied at the density functional theory (DFT) level using the Quantum Espresso[7] package. The exchange-correlation functional is approximated using the generalized gradient approximation[8]. We use norm-conserving pseudopotentials[9,10] and a vacuum of 22 Å is included in the supercell calculation to minimize interaction between periodic slabs in the out-of-plane direction. The wavefunctions are expanded in planewaves up to an energy cutoff of 40 Ry. The moiré Brillouin zone (BZ) is sampled only at the γ point to obtain the self-consistent potential and charge density. Spin-orbit coupling is not explicitly included in our DFT calculations.

### S1.2 Excited state calculations

The DFT band structure captures the relative dispersion of the quasiparticle Bloch states close to the valence band edge and conduction band edge in TMDs[13], but severely underestimates the band gap. We compute the quasiparticle correction (using the *GW* approximation[14]) to the DFT band gap for the pristine unit-cell $WSe_2$ and use the same correction (including spin-orbit coupling effects) for the band gap of the reconstructed $WSe_2$. A one-shot *GW* calculation is carried out using the BerkeleyGW[15] package. The static dielectric matrix, computed using the random phase approximation with a planewave energy cut-off of 30 Ry and ~6000 unoccupied states, is extended to finite frequencies using the Hybertsen-Louie generalized plasmon pole model[14]. The nonuniform neck subsampling[16] method is used to improve convergence of the band gap with k-point sampling of the BZ.

To obtain the optical spectrum of the strained $WSe_2$ superlattice, we first compute the electron-hole singlet kernel matrix elements for the pristine unit-cell $WSe_2$: $\langle \phi^{cond}_{\alpha k_{uc}+Q} \phi^{val}_{\beta k_{uc}} | K^p | \phi^{cond}_{\gamma k'_{uc}+Q} \phi^{val}_{\eta k'_{uc}} \rangle$, where the states $\phi$ are the pristine unit-cell wavefunctions for a band and k-point in the unit-cell BZ, $Q$ is the exciton center-of-mass wavevector in the unit-cell BZ. These kernel matrix elements along with the wavefunction expansion coefficients ($a_i$) are used to obtain the kernel matrix elements of the strained $WSe_2$ superlattice as described in the main text. Since this evaluation is a coherent sum, the gauge of the wavefunctions used to compute the unit-cell kernel matrix elements must match the gauge of the pristine (i.e., unstrained) supercell $WSe_2$ wavefunctions (denoted by Φ) used to compute the expansion coefficients. To ensure a consistent gauge, we construct the pristine supercell wavefunction from the corresponding pristine unit-cell state:

$$\Phi_{i\boldsymbol{k}_{sc}} = \sum_{\boldsymbol{G}_{sc}} c_{\alpha \boldsymbol{k}_{uc}}(\boldsymbol{G}_{uc}) e^{i(\boldsymbol{k}_{uc}+\boldsymbol{G}_{uc}).\boldsymbol{r}} \delta_{\boldsymbol{k}_{uc}+\boldsymbol{G}_{uc},\boldsymbol{k}_{sc}+\boldsymbol{G}_{sc}}, \quad (S1)$$

where the band $i$ and k-point $\boldsymbol{k}_{sc}$ of the supercell is related to the band $\alpha$ and k-point $\boldsymbol{k}_{uc}$ in the unit-cell by band folding. $c_{\alpha \boldsymbol{k}_{uc}}(\boldsymbol{G}_{uc})$ are the planewave coefficients of the unit-cell wavefunction. The exciton spectrum of the strained WSe$_2$ superlattice is obtained by solving the Bethe-Salpeter equation (BSE):

$$(E_{c\boldsymbol{k}} - E_{v\boldsymbol{k}})A^S_{vc\boldsymbol{k}} + \sum_{v',c',\boldsymbol{k}'} \langle vc\boldsymbol{k}|K^S|v'c'\boldsymbol{k}'\rangle A^S_{v'c'\boldsymbol{k}'} = \Omega_S A^S_{vc\boldsymbol{k}}, \quad (S2)$$

where $E_{c\boldsymbol{k}}$ and $E_{v\boldsymbol{k}}$ are the quasiparticle eigenvalues corresponding to a conduction and valence state in the moiré BZ, respectively. $\Omega_S$ and $A^S_{vc\boldsymbol{k}}$ are the exciton eigenvalues and eigenvectors, respectively. We use a k-point sampling of $3 \times 3 \times 1$ in the moiré BZ, 12 valence and 12 conduction bands to construct the BSE Hamiltonian. We find that a finer k-point sampling of $6 \times 6 \times 1$, does not change the absorption spectrum significantly (Fig. S1). The influence of spin-orbit coupling is the formation of two nearly identical series of excitonic peaks (shifted in energy) in the absorption spectrum: the A and B series. A recent study show that the intrinsic spin-orbit splitting of the band states derived from the $K$ point in the unit-cell BZ is unaffected by structural reconstructions in the moiré superlattice[11]. We account for spin-orbit coupling as a perturbation to our absorption spectrum[12], $\Delta\Omega^S_\sigma = \sum_{vc\boldsymbol{k}}|A^S_{vc\boldsymbol{k}}|^2 \Delta\epsilon^{SO}_{vc\boldsymbol{k}\sigma}$, where the $\Delta\epsilon^{SO}_{vc\boldsymbol{k}\sigma}$ are the corrections to the valence and conduction eigenvalues (with spin $\sigma$) due to the inclusion of spin-orbit coupling[12]. We find that, while the rigid shift in the eigenvalues due to the spin-orbit splitting is large, the change in energy dispersion of the electronic states is relatively small (less than 10 meV), i.e., $\Delta\Omega^S_\sigma \approx \Delta\epsilon^{SO}_\sigma \sum_{vc\boldsymbol{k}}|A^S_{vc\boldsymbol{k}}|^2$. Furthermore, since the kernel matrix-elements are computed in a spin-singlet basis, it leads to a double counting of the like-spin transitions. This leads to a doubling of the oscillator strength for each of the exciton peaks when compared to a calculation including spin-orbit coupling. Hence, in addition to correcting the exciton eigenvalues, we also accordingly renormalize our absorbance spectra with a factor of 0.5.

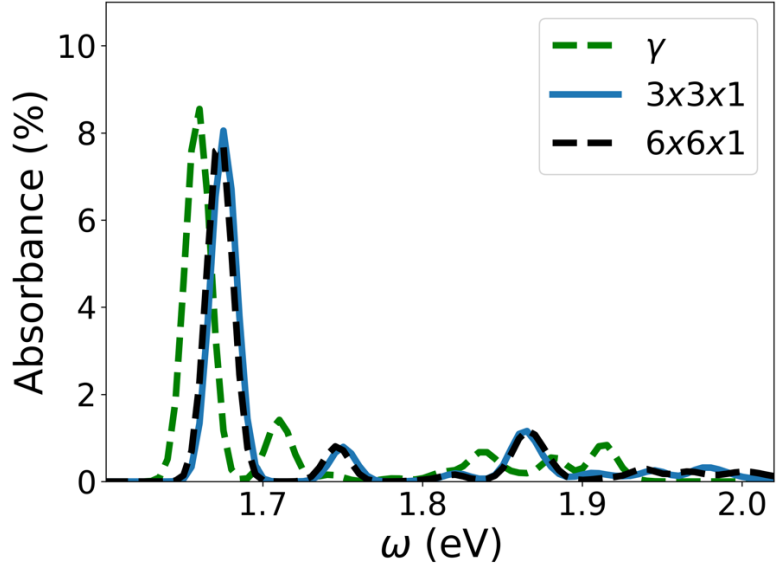

**Fig. S1. | Convergence of the calculated absorption spectrum with different k-point samplings of the moiré Brillouin zone.** γ refers to a sampling of only the zone center γ point of the moiré BZ.

**S2. Influence of structural reconstruction on the electronic band structure**

The bilayer moiré superlattice is composed of various local stacking configurations which have different structural energies. The stackings with AA configuration have Se of $WSe_2$ facing S of $WS_2$ which makes them energetically less stable than the Bernal stackings[17,18]. This leads to a reconstruction of the superlattice in the form of a strain redistribution in each layer as well as an out-of-plane undulation of the superlattice. These reconstructions in rotationally aligned $WSe_2/WS_2$ and their influence on the electronic structure has been recently studied using scanning tunneling microscopy combined with theoretical calculations[19].

We validate the various approximations made in the main text by performing a large scale DFT calculation of the entire moiré superlattice. Since the moiré unit-cell contains a large number of atoms (3903), studying the electronic structure of the full system at the DFT level is computationally challenging using the planewave basis set. This is because the planewave DFT calculations typically scale as $N^2 \log(N)$ with the number of atoms. Hence, for the purpose of validating the approximations used, we use a computationally more tractable DFT scheme based on a basis set of atomic orbitals as implemented in the SIESTA package[20]. The calculations are performed using norm-conserving[21] pseudopotentials, a real-space grid equivalent to a plane-wave energy cut-off of 80 Ry is used, and only the γ point is sampled in the moiré BZ. Spin-orbit coupling is included in these calculations and the local density approximation is used for the exchange-correlation functional. Note that the calculations presented in the main text, where we study the electronic and optical properties of the strained $WSe_2$ superlattice, have been carried out using planewave DFT calculations (computed using the Quantum Espresso[7] package).

The WSe$_2$/WS$_2$ heterostructure forms a type II heterojunction with the valence band edge derived from the WSe$_2$ layer and the conduction band edge from the WS$_2$ layer (Fig. S2a). The electronic structure of the reconstructed bilayer shows energy-separated flat moiré bands at the valence and conduction band-edge complexes. In contrast, a rigidly aligned bilayer (at an average interlayer spacing) does not show these flat band features. Including structural reconstructions in the calculation is thus critical to obtain the right electronic structure of the moiré superlattice. Furthermore, we find that the valence band states of the rigidly aligned bilayer and the reconstructed bilayer closely resemble those of the pristine monolayer WSe$_2$ and reconstructed monolayer WSe$_2$, respectively. This indicates that these moiré states (derived from the WSe$_2$ $K$ point in the unit-cell BZ) are not strongly hybridized with the WS$_2$ layer in the bilayer and justifies the approximation of including only the reconstructed WSe$_2$ layer in our BSE calculations.

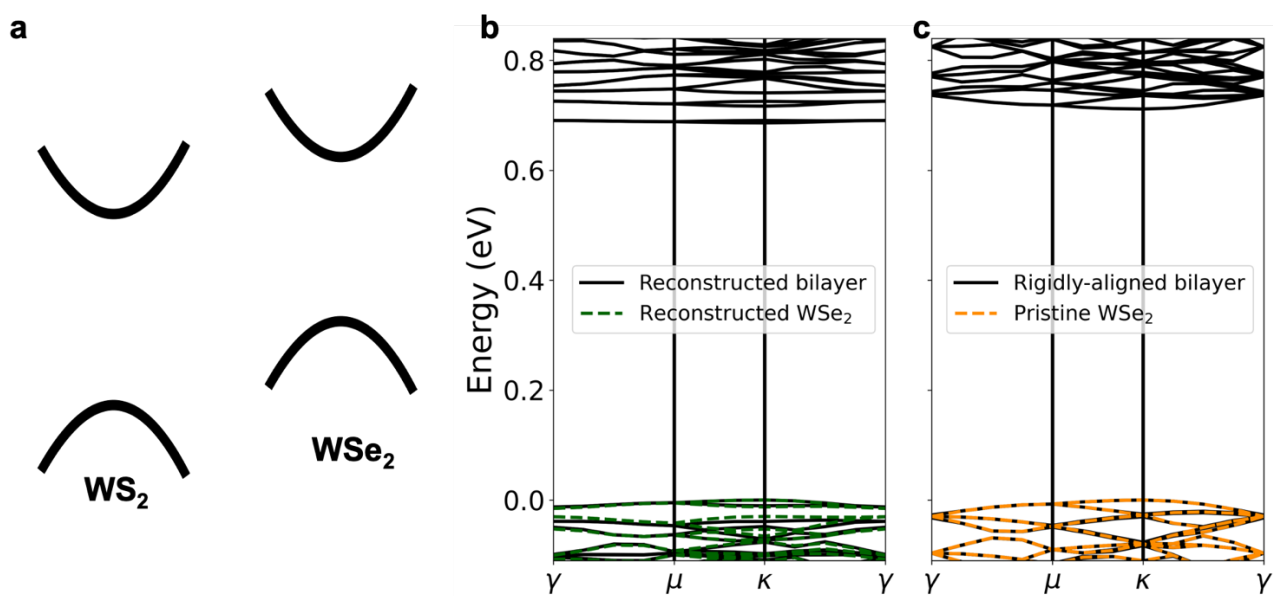

**Fig. S2. | Electronic structure of reconstructed and rigidly aligned WSe$_2$/WS$_2$ heterostructure. (a)** Type II band alignment of the $K$ valley band states in the WSe$_2$/WS$_2$ heterostructure. The valence band-edge states in the bilayer correspond to the WSe$_2$ states, while the conduction band-edge states correspond to WS$_2$. (**b** and **c**) Comparison of electronic band structure of the reconstructed and rigidly aligned bilayers. The reconstructed superlattice has flatter bands and larger gap openings at the Brillouin zone edges. The band structure of the rigidly aligned bilayer, on the other hand, virtually identical to the folded band structure of the pristine monolayer WSe$_2$ (orange-dashed line) at the valence band edge. Furthermore, the electronic band structure of the reconstructed monolayer WSe$_2$ is very similar to the bilayer valence states (green-dashed line).

We make an additional approximation in our simulation of the reconstructed WSe$_2$ superlattice. We remove the out-of-plane reconstruction of the WSe$_2$ layer by changing the height of each atom to the pristine value (Fig. S3a), while keeping all in-plane reconstructions intact. This is reasonable since the undulations are smooth and vary slowly over the large moiré length scale. The additional strain induced due to this transformation is ~0.05%, which is negligible compared to the large strains (~1%) induced by the in-plane reconstruction. The out-of-plane reconstructions, while important to stabilize the structure of the bilayer moiré superlattice, have an insignificant impact on the electronic structure (Fig. S3b). This transformation helps with the construction of electron-hole interaction kernel in the strained $WSe_2$ superlattice in terms of pristine unit-cell kernel matrix elements. The small effects of the out-of-plane corrugations also justify our use of the screened Coulomb interaction of the strained $WSe_2$ superlattice being that of pristine $WSe_2$.

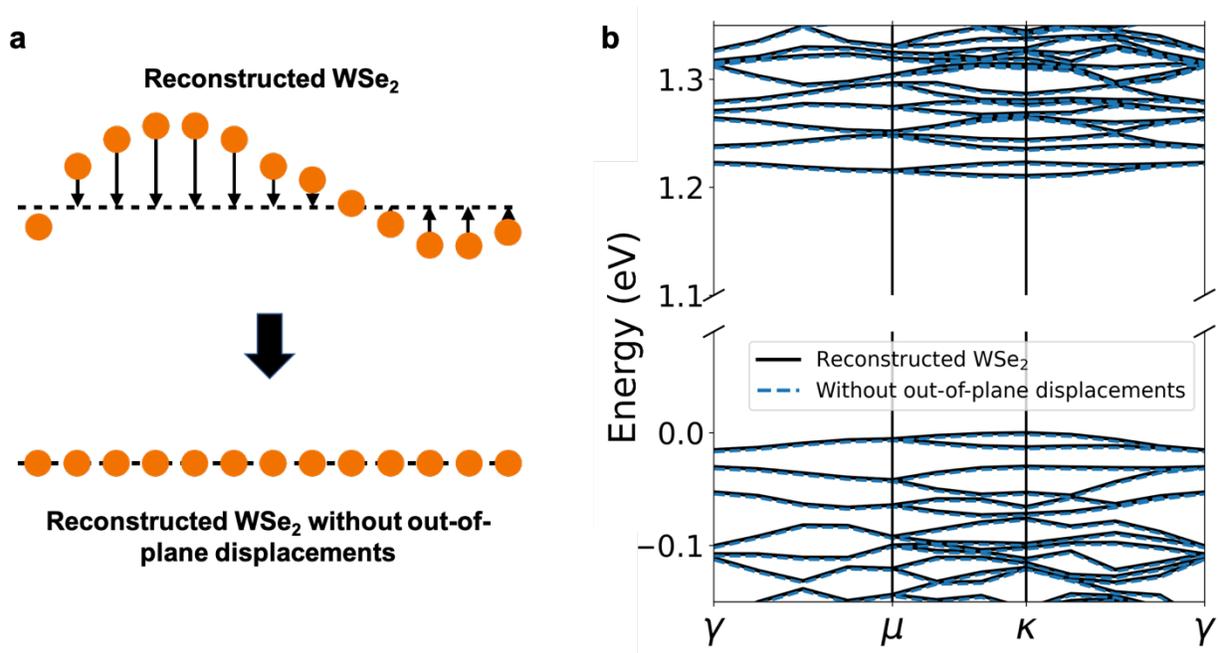

**Fig. S3. Influence of out-of-plane displacements to the electronic structure.** (**a**) Schematic of transformation (the vertical displacement is grossly exaggerated by a factor of 100 for illustration) of the reconstructed $WSe_2$ layer to remove out-of-plane displacements while retaining the in-plane reconstruction. The position of atoms is kept intact in the in-plane direction while vertical displacements from the monolayer configuration are set to zero. (**b**) Comparison of the electronic band structure of the reconstructed $WSe_2$ with and without out-of-plane displacements.

## S3. Pristine unit-cell Matrix Projection (PUMP) method

The pristine unit-cell matrix projection (PUMP) method introduces a basis transformation to make the computation of the electron-hole interaction kernel of the BSE tractable. We first express the single-particle electron ($\psi_{c\mathbf{k}_m}$) and hole wavefunctions ($\psi_{v\mathbf{k}_m}$) of the moiré superlattice as a coherent superposition of pristine unit-cell wavefunctions: $|\psi_{v\mathbf{k}_m}\rangle = \sum_i a_i^{v\mathbf{k}_m} |\Phi_{i\mathbf{k}_m}^{val}\rangle$, $|\psi_{c\mathbf{k}_m}\rangle = \sum_i a_i^{c\mathbf{k}_m} |\Phi_{i\mathbf{k}_m}^{cond}\rangle$. The states $|\Phi_{i\mathbf{k}_m}\rangle$ here refer to the pristine supercell wavefunctions that are constructed from the corresponding unit-cell wavefunctions (related by band-folding) while preserving the gauge, as described in section S1. The pristine supercell wavefunctions have a phase winding in the moiré superlattice corresponding to the unit-cell BZ k-point that the state unfolds to (Fig. S4a). The constructed moiré wavefunctions are in excellent agreement with the original wavefunctions, as shown in Fig. S4 b and c. Here, we expressed 12 valence and 12 conduction moiré wavefunctions at a given k-point in the moiré BZ in a basis of 24 valence and 24 conduction pristine (e.g, band-folded) states. The projections of the constructed valence and conduction states on the original moiré wavefunctions are shown in Fig. S4d. The large projection (>95%) confirms that the chosen small pristine unit-cell basis is suitable to expand the moiré electronic wavefunctions. The expansion coefficients, $a_i^{v\mathbf{k}_m}$ and $a_i^{c\mathbf{k}_m}$, are subsequently used to expand the moiré kernel matrix elements in terms of unit-cell kernel matrix elements:

$$\langle \psi_{c\mathbf{k}_m} \psi_{v\mathbf{k}_m} | K^s | \psi_{c'\mathbf{k}'_m} \psi_{v'\mathbf{k}'_m} \rangle$$

$$\approx \sum_{i,j,p,q} a_i^{c\mathbf{k}_m*} a_j^{v\mathbf{k}_m*} a_p^{c'\mathbf{k}'_m} a_q^{v'\mathbf{k}'_m} \left\langle \Phi_{i\mathbf{k}_m}^{cond} \Phi_{j\mathbf{k}_m}^{val} | K^p | \Phi_{p\mathbf{k}'_m}^{cond} \Phi_{q\mathbf{k}'_m}^{val} \right\rangle \tag{S3}$$

$$= \sum_{i,j,p,q} a_i^{c\mathbf{k}_m*} a_j^{v\mathbf{k}_m*} a_p^{c'\mathbf{k}'_m} a_q^{v'\mathbf{k}'_m} \left\langle \phi_{\alpha\mathbf{k}_{uc}^1}^{cond} \phi_{\beta\mathbf{k}_{uc}^2}^{val} | K^p | \phi_{\gamma\mathbf{k}'^1_{uc}}^{cond} \phi_{\eta\mathbf{k}'^2_{uc}}^{val} \right\rangle \delta_{\mathbf{k}'^2_{uc}-\mathbf{k}'^1_{uc}, \mathbf{k}^2_{uc}-\mathbf{k}^1_{uc}}, \tag{S4}$$

where the pristine states corresponding to $i\mathbf{k}_m, j\mathbf{k}_m, p\mathbf{k}'_m, q\mathbf{k}'_m$ in the moiré BZ map to $\alpha\mathbf{k}_{uc}^1, \beta\mathbf{k}_{uc}^2, \gamma\mathbf{k}'^1_{uc}, \eta\mathbf{k}'^2_{uc}$ states in the unit-cell BZ (Fig. S4e).

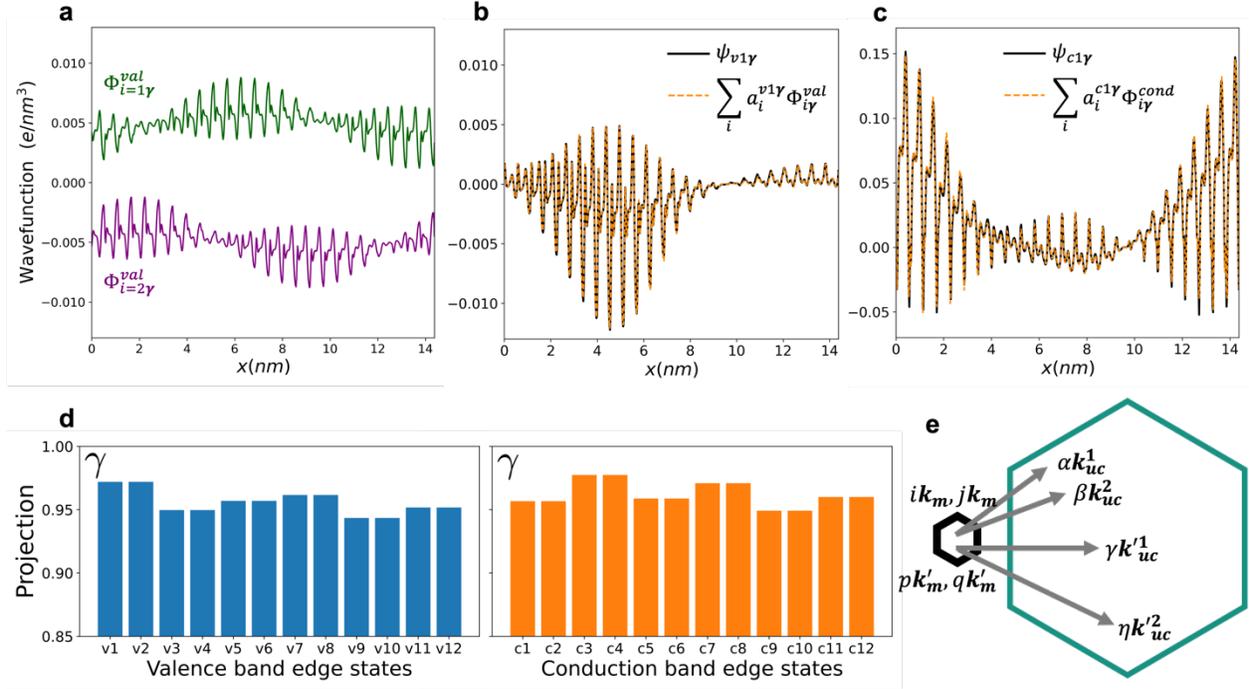

**Fig. S4. Pristine unit-cell expansion of moiré wavefunctions.** (**a**) Two representative Bloch state of the pristine WSe$_2$ crystal in real space, arbitrarily displaced along the vertical axis for illustration. The wavefunction is averaged along the out-of-plane direction and the real part is plotted along the (110) direction in the pristine superlattice. (**b**) Real part of the strained-superlattice wavefunction (black solid line) corresponding to the valence band maximum at the γ point. This is compared to the wavefunction constructed (orange dashed line) as a linear superposition of the pristine basis states. $x = 0$ here corresponds to the AA stacking in the superlattice. (**c**) Similar plot as (**b**) comparing the conduction band minimum at the γ point. (**d**) Projection (sum of the square of the overlaps) of the strained superlattice valence and conduction wavefunctions, at the γ point in the moiré BZ, onto a basis of 24 pristine valence and 24 pristine conduction states, respectively. (**e**) The mapping of states involved in a kernel matrix element (described in the text) from the supercell BZ (black hexagon) to the unit-cell BZ (green hexagon).

## S4. Validation of the pristine unit-cell matrix projection (PUMP) method

The electron-hole interaction kernel of the Bethe-Salpeter equation is composed of two terms: the direct and the exchange term. The direct term, which accounts for the attractive screened Coulomb interaction between the electron and hole is

$$\langle \psi_{ck}\psi_{vk}|K_d|\psi_{c'k'}\psi_{v'k'}\rangle = -\int d\mathbf{r}\, d\mathbf{r}'\, \psi_{ck}^*(\mathbf{r})\psi_{c'k'}(\mathbf{r})W(\mathbf{r},\mathbf{r}')\psi_{v'k'}^*(\mathbf{r}')\psi_{vk}(\mathbf{r}'), \quad (S5)$$

where $W$ is the static screened Coulomb interaction. We use the PUMP technique to express the kernel of the strained WSe$_2$ superlattice in terms of pristine unit-cell WSe$_2$ kernel matrix elements. Here, we assume that the screened Coulomb interaction of the strained superlattice is identical to that of pristine WSe$_2$. This approximation is robust since the excitons in TMDs have a lateral size of ~2 nm. Hence, the screening is not sensitive to small changes in the charge-density at the atomic scale. We validate this approximation by explicitly solving the BSE for a smaller, computationally tractable superlattice. We construct a smaller "model" moiré superlattice using a 19 × 19 supercell of $WSe_2$ and a 20 × 20 supercell of $WS_2$. We strain the lattice parameters of $WS_2$ by ~1.1% here so that the supercell matching condition is satisfied, and we have a periodic superlattice. On relaxing this system, we find a similar strain network with a maximum tensile and compressive strains of 0.8% and -0.6%. These strains are smaller than the 25 × 25 superlattice used in our main theoretical study. We compare absorption spectrum of the 19x19 strained $WSe_2$ superlattice computed explicitly to the spectrum calculated using Eq. (S4) above for the electron-hole kernel in Fig. S5. The two spectra are in excellent agreement, and the two sets of exciton eigenvalues have a deviation of only ~5 meV. This validates our assumption and clearly shows that the PUMP method developed can capture the moiré physics of the TMD hetero-bilayers accurately. To ease the computational cost associated with computing the superlattice kernel directly, we included 6 valence, 6 conduction bands and sample only the $\gamma$ point in the moiré BZ for these validation calculations. A basis of 12 pristine valence and 12 pristine conduction bands (in the pristine $WSe_2$ supercell) is used to expand the wavefunctions of the strained $WSe_2$. Only two distinguishable peaks are seen in the absorption spectrum here due to the relatively smaller strains in the model superlattice, and more importantly, lack of fine k-point sampling of the moiré BZ.

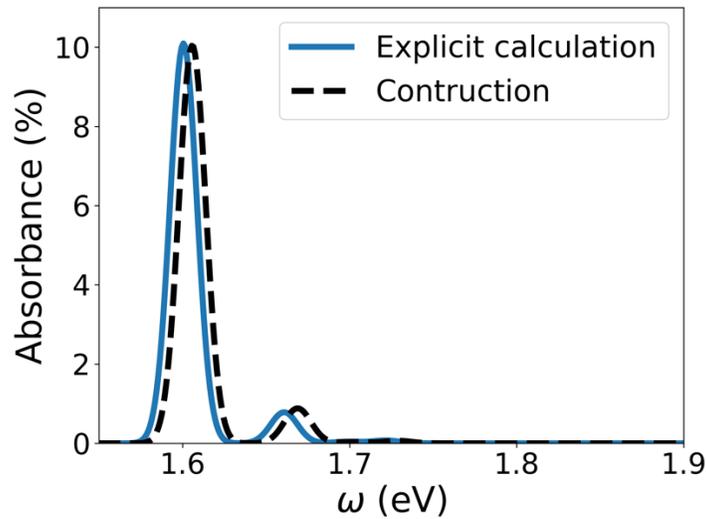

**Fig. S5. Validation of the pristine unit-cell matrix projection (PUMP) method.** Comparison of the explicitly calculated absorption spectrum of a reconstructed 19x19 $WSe_2$ superlattice with the spectrum constructed using the pristine unit-cell matrix projection (PUMP) method.

## S5. Analysis of the moiré exciton peaks

Figure S6 shows the band composition of the calculated moiré exciton wavefunction ($A^S_{vc\boldsymbol{k}_m}$), obtained by integrating out the k-points, $\sum_{\boldsymbol{k}_m}|A^S_{vc\boldsymbol{k}_m}|^2$. Moiré exciton forming peak I is composed of transitions involving multiple moiré valence and primarily the two lowest moiré conduction bands (c1 and c2), because the strong electron-hole Coulomb attraction mixes different electron-hole states from different bands effectively. Exciton II is also spread over multiple moiré valence and conduction states. The highest-energy moiré exciton III has contributions dominated by the highest valence and lowest conduction band edge states, consistent with its smaller binding energy.

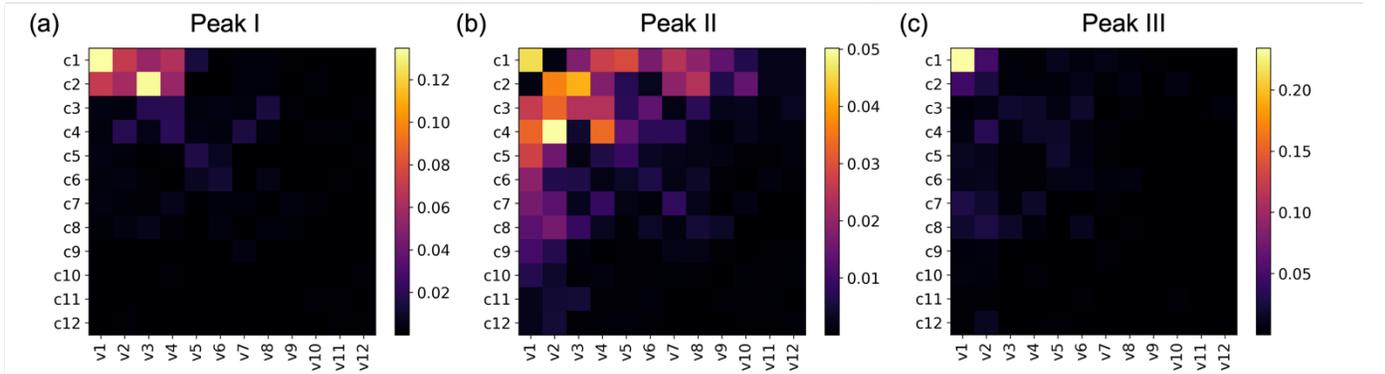

**Fig. S6. Decomposition of exciton amplitude into valence and conduction band contributions.** Contribution of the moiré valence and conduction band states, as defined in the SI text, to the exciton corresponding to the three strong moiré exciton resonances.

Furthermore, we find that the interband velocity matrix element, $\langle v\boldsymbol{k}_m|\mathbf{v}|c\boldsymbol{k}_m\rangle$, are significantly altered by the modulation of the electronic wavefunctions (Fig. S7). For a pristine superlattice (that is, the WSe$_2$ crystal's unit cell is arbitrarily enlarged from the primitive unit cell to a non-primitive supercell with large area), the velocity matrix elements are determined by their corresponding band states in the primitive unit-cell BZ, i.e. $\langle v\boldsymbol{k}_m|\mathbf{v}|c\boldsymbol{k}_m\rangle = \langle \alpha\boldsymbol{k}^1_{uc}|\mathbf{v}|\beta\boldsymbol{k}^2_{uc}\rangle \delta_{\boldsymbol{k}^1_{uc},\boldsymbol{k}^2_{uc}}$ where the states $v\boldsymbol{k}_m$ and $c\boldsymbol{k}_m$ in the moiré BZ map to the states $\alpha\boldsymbol{k}^1_{uc}$ and $\beta\boldsymbol{k}^2_{uc}$ in the primitive unit-cell BZ through band folding, respectively. Crystal momentum indirect transitions are forbidden in a pristine layer. The moiré wavefunctions of a strained WSe$_2$, on the other hand, can be a mixture of several (with different *k and band indices*) pristine Bloch band states, causing a qualitative change in the value of these matrix elements. The interband transition matrix elements in the strained superlattice are strongest between moiré band states with large wavefunction amplitude in same spatial region in the moiré superlattice. For example, the substantial wavefunction overlap between states in bands v3 and c2 (Fig. 1e and f in the main text) leads to the large oscillator strength observed for peak I in the absorbance spectrum. The relatively smaller velocity matrix elements from v1 to c1 around the $\kappa$ point in the moiré BZ leads to the smaller oscillator strength of peak III.

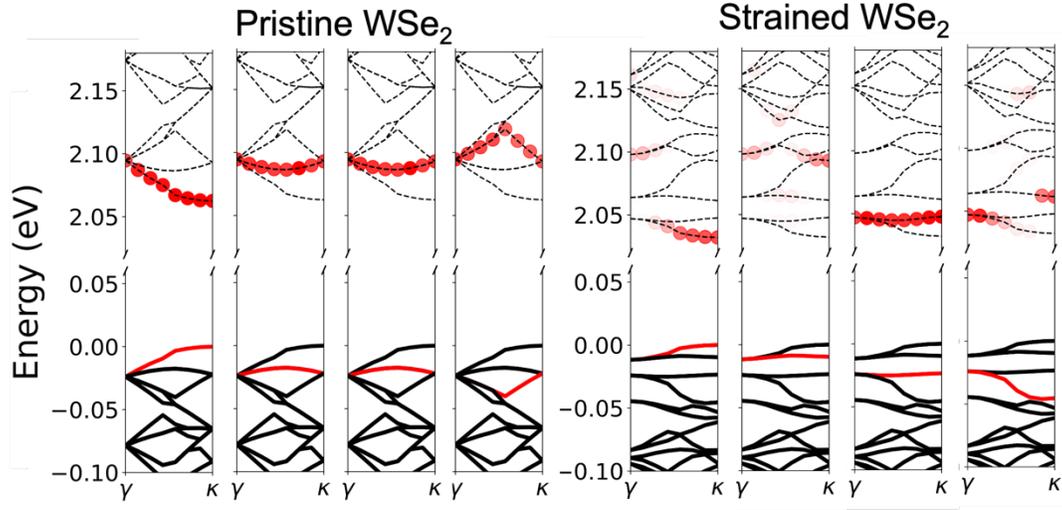

**Fig. S7. | Value of interband optical transition matrix elements from a valence band (marked in red) to all the conduction bands.** The transparency of the red dots in the conduction band indicates the relative strength of the matrix elements, with the darker color representing larger matrix element. For the pristine WSe$_2$ layer with bands folded to the corresponding moiré BZ, only matrix elements of transitions that correspond to momentum-direct transitions in the unit-cell Brillouin zone are nonzero. The inhomogeneous distribution of strains in the strained WSe$_2$ superlattice (a real moiré crystal) changes the character of the wavefunctions of states at the moiré valence and moiré conduction bands leading to a qualitative change in the transition matrix elements.

An alternative approach to characterize the spatial modulation of the moiré excitons is by computing the hole (electron) charge density that contributes to the exciton by integrating out the electron (hole) coordinate from the exciton wavefunction: $\int d\boldsymbol{r_e}|\chi_S(\boldsymbol{r_e},\boldsymbol{r_h})|^2$ ($\int d\boldsymbol{r_h}|\chi_S(\boldsymbol{r_e},\boldsymbol{r_h})|^2$) where the integration over the electron (hole) position is over all space. For the case of an exciton in a pristine layer, these charge densities would be uniformly distributed across the layer with no preference for any particular sites. For the strained moiré system, the hole and electron density is maximum at the AA stacking site for exciton I (Fig. S8). This exciton thus possesses a modulated Wannier character as described in the main text. Peak II exciton has a partially Wannier character with some regions of hole and electron overlap. Peak III exciton, in contrast, has the hole and electron densities clearly separated in the moiré superlattice with the electron density maximum at the AA site and hole density at the B$^{Se/W}$ stacking site. Peak III thus forms a charge-transfer exciton.

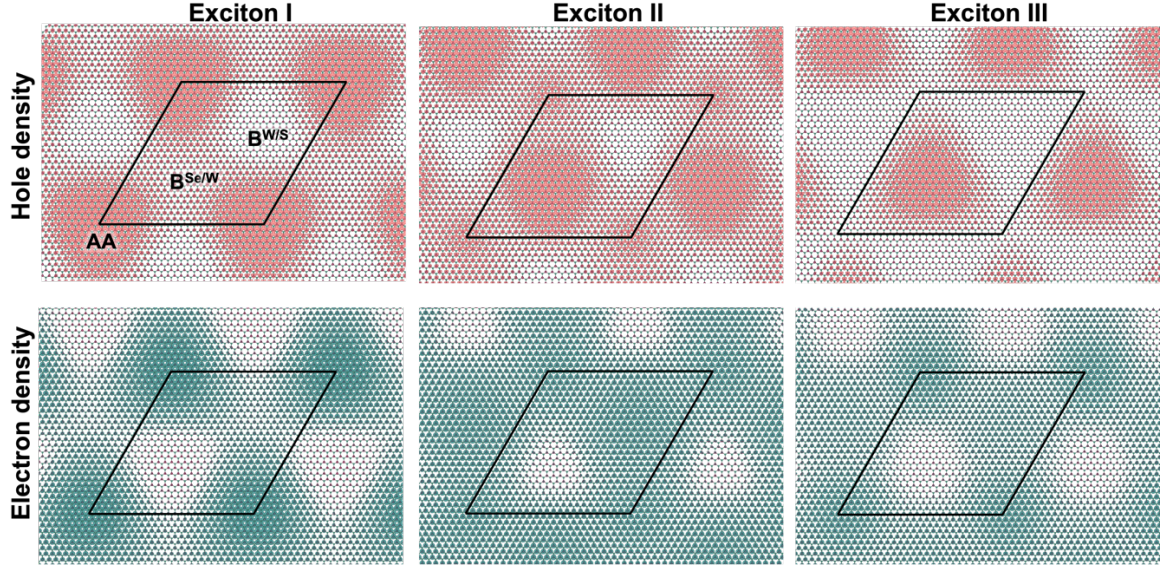

**Fig. S8. | Hole (top) and electron (bottom) charge densities for the exciton states corresponding to the moiré exciton peaks I, II, and III.** The hole (electron) charge density is computed by integrating out the electron coordinates (hole coordinates) from the squared exciton amplitude over all space, as defined in the SI text.

### S6. Continuum model for the moiré excitons

As described in the main text, the continuum model[22] is based on perturbing the intralayer A exciton with an effective moiré potential ($\Delta(\mathbf{R})$):

$$H = H_0 + \Delta(\mathbf{R}) \tag{S6}$$

where $H_0$ is the low-energy, effective two-particle Hamiltonian corresponding to the WSe$_2$ A (1s) exciton and $\mathbf{R}$ is the exciton center-of-mass coordinate. $H_0$, in momentum space, describes the exciton band structure for transitions around the K point in the unit-cell BZ[22,23].

$$H_0(\mathbf{Q}) = \left(\Omega_0 + \frac{\hbar^2 \mathbf{Q}^2}{2M}\right)\sigma_0 + J|\mathbf{Q}|[\sigma_0 + \cos(2\phi_\mathbf{Q})\sigma_x + \sin(2\phi_\mathbf{Q})\sigma_y] \tag{S7}$$

where $\mathbf{Q}$ is the exciton center-of-mass momentum in the unit-cell BZ; $\sigma_0$ is the 2x2 identity matrix; $\sigma_x$ and $\sigma_y$ are Pauli matrices; $\Omega_0$ is the excitation energy of the $\mathbf{Q} = \mathbf{0}$ exciton. M and J are parameters that can be fit to the A exciton dispersion relation computed from ab initio. The values of M and J used in the present model are 1.0 m$_e$ and 0.8 eVÅ. The effective potential for the excitons is modelled as the variation in band gap as a function of the local strain in the superlattice (Fig. 1 of main text). The local strain at a given W atom site in the superlattice is evaluated by averaging the strain along six directions corresponding to the six neighboring W atoms. A unit-cell DFT calculation is performed incorporating this strain to evaluate the change in bandgap. The effective potential is approximated as a smooth potential[22], $\Delta(\mathbf{R}) = \sum_{j=1}^{6} V_j e^{i\mathbf{b}_j \cdot \mathbf{R}}$ with $V_1 = V_3 = V_5 = V e^{i\psi}$; $V_2 = V_4 = V_6 = V e^{-i\psi}$; and $\mathbf{b}_j$ are the reciprocal lattice vectors. The value of $(V, \psi)$ that fits the computed band gap variation is (9.63 meV, -175.1°).

## S7. Estimation of electron doping dependence of peak III

As described in the main text, upon electron doping, our measurements show that peak III red-shifts by about a 100 meV. To obtain a rough estimation of this shift, we compute the electrostatic interaction between the doping electron and charge-transfer exciton, using our calculated results. First, we compute the screened Coulomb interaction strength of the doping electron and the constituent hole of the exciton, which both have their maximum density at the $B^{Se/W}$ site. The two charges are separated by 0.7 nm in the out-of-plane direction. Considering the computed charge density distribution of the hole and the electron and assuming a uniform dielectric constant of $\epsilon = 4$ from the TMD and hBN layers, we obtain an electrostatic energy for this interaction to be: $\delta E_{e-h} \approx -170\ meV$. Next, we compute the electrostatic interaction between the doping electron (at the $B^{Se/W}$) and the constituent electron of the exciton (at the AA site). This distance is much larger, $\sim 5.4\ nm$, hence we can consider it as point charges interacting. The resulting repulsive energy is, $\delta E_{e-e} \approx 50\ meV$. The total electrostatic energy of the interaction between the doping electron and exciton in this simple model is $\sim -120\ meV$. This is of course only an order-of-magnitude estimate since we do not include the anisotropy of the dielectric environment and local-field effects in the calculation.

## S8. Device information

We first exfoliate WS$_2$ and WSe$_2$ monolayers from bulk crystal onto SiO$_2$/Si substrate and then do polarization-resolved second harmonic generation (SHG) measurements to determine the crystal orientation[24,25]. We subsequently pick up WS$_2$ and WSe$_2$ flakes using a polypropylene carbon (PPC) or polyethylene terephthalate (PET) stamp[26]. The near zero-degree-twist-angle WSe$_2$/WS$_2$ heterostructure is contacted by few-layer graphite (FLG) and sandwiched by the two hexagonal boron nitride (hBN) dielectric layers. Additional FLG flakes serve as top and/or bottom gates. The whole heterostructure is released on a 90 nm SiO$_2$/Si substrate. The electrodes (5 nm Cr /100 nm Au) are fabricated by a standard photolithography system (Durham Magneto Optics, MicroWriter) and an e-beam deposition system. After fabrication, we again perform polarization-resolved SHG measurements on the monolayer regions of the sample to determine the exact twist angle and on the heterostructure region to distinguish between near-zero and near-sixty-degree samples.

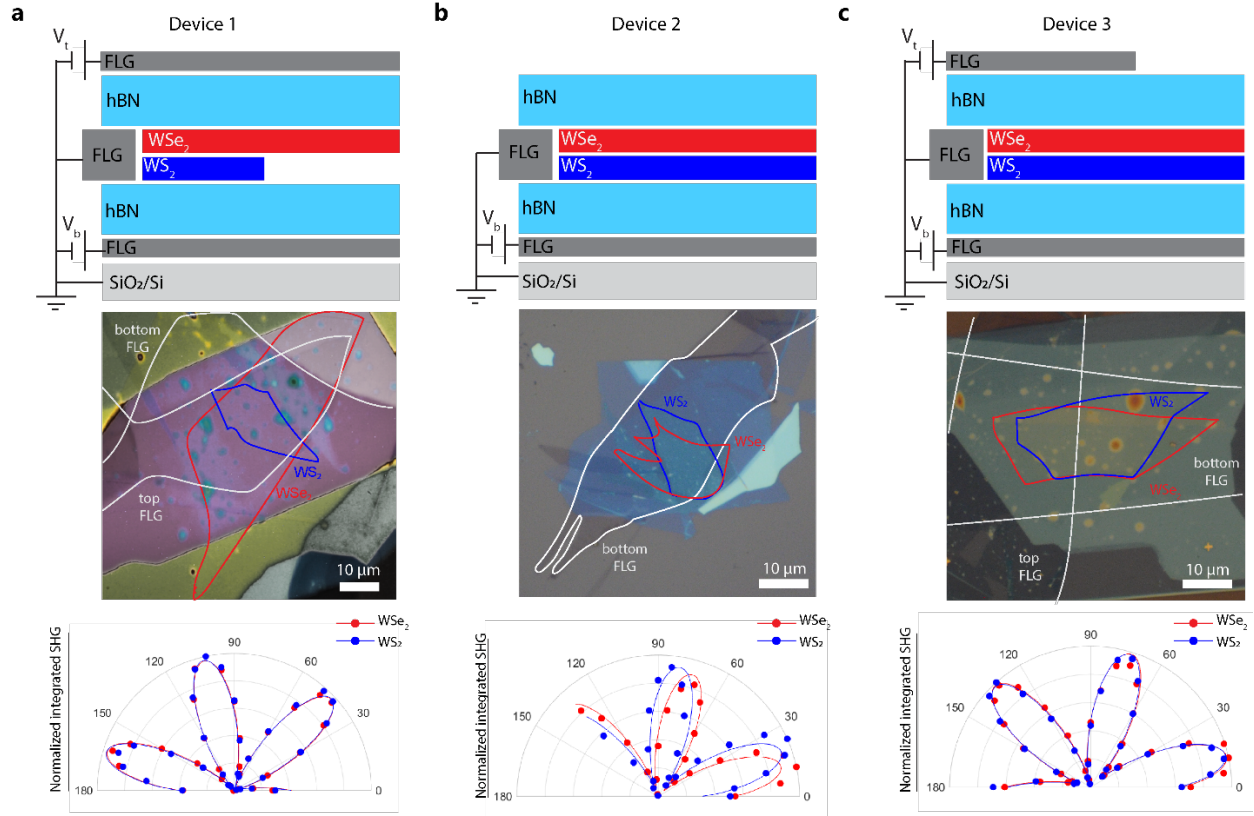

**Fig. S9 | Device information.** Sample schematics (upper panel), false-color optical microscope images (middle panel), and polarization-resolved SHG data (bottom panel) for three devices included in this study. In SHG data, the red (blue) circles correspond to normalized integrated SHG at different polarization angles θ for the $WSe_2$ ($WS_2$) layers, and the solid lines are corresponding fits to a $\cos^2(3\theta)$ function. (**a**) Device 1: Near-zero-degree, dual-gated $WSe_2/WS_2$ heterostructure described in main text. $WSe_2$ monolayer region also described in main text. (**b**) Device 2: Misaligned $WSe_2/WS_2$ heterostructure (6 degrees) described in main text. (**c**) Device 3: Additional near-zero-degree, dual-gated $WSe_2/WS_2$ heterostructure described in S13

## S9. Doping and magnetic field dependent reflection contrast measurements

The micro reflection contrast measurements are performed in a cryostat at a temperature of T = 1.6 K and a magnetic field of up to B = 6 T (Quantum Design, Opticool). A diode laser is focused on the sample by a 20X Mitutoyo objective with ~2 μm beam size. The diode is driven below the lasing threshold, producing a relatively broadband beam centered around 730 nm. The reflection contrast is defined as the normalized difference in reflection on the sample and the substrate (including all hBN and FLG layers, but no TMD layers). The spectra include a slowly varying background due to multi-layer interference, so a polynomial background is fitted to the data and subtracted to observe the exciton states more easily[27].

Doping dependent measurements are conducted by applying voltages to the top and bottom gates using Keithley 2400 or 2450 source meters. The charge density is defined using a parallel plate capacitor model:

$$n = \pm\frac{1}{e}\left[\frac{\epsilon_{hBN}\epsilon_0}{d_t}(V_t - V_{t0}) + \frac{\epsilon_{hBN}\epsilon_0}{d_b}(V_b - V_{b0})\right],$$

where $\epsilon_{hBN}$ is the dielectric constant of hBN (measured as $4.2 \pm 0.2$ in Ref. [28]), $\epsilon_0$ is the permittivity of free space, $d_t$ ($d_b$) is the thickness of the top (bottom) hBN layer, and $V_t$ ($V_b$) is the voltage applied to the top (bottom) FLG gate. We account for the quantum capacitance (voltage range where charges are not injected) using $V_{t0}$ and $V_{b0}$, offset voltages defined where the spectrum begins to change with electron and hole doping, respectively. The vertical electric field is set to 0 V/nm in the reflection contrast measurements of the moiré superlattice shown in the main text. In the case of the large-twist-angle sample, device 2, the sample only has a bottom gate, so the electric field is finite but does not strongly affect the WSe$_2$ A exciton due to the lack of out-of-plate dipole.

For reflection contrast measurements in the magnetic field, a quarter-wave plate is placed above the objective to convert the linearly polarized light to circularly polarized light. The reflected light, collected by the same objective, goes through the same quarter-wave plate before reaching the camera.

## S10. Doping and magnetic field dependent reflection contrast data at higher charge doping

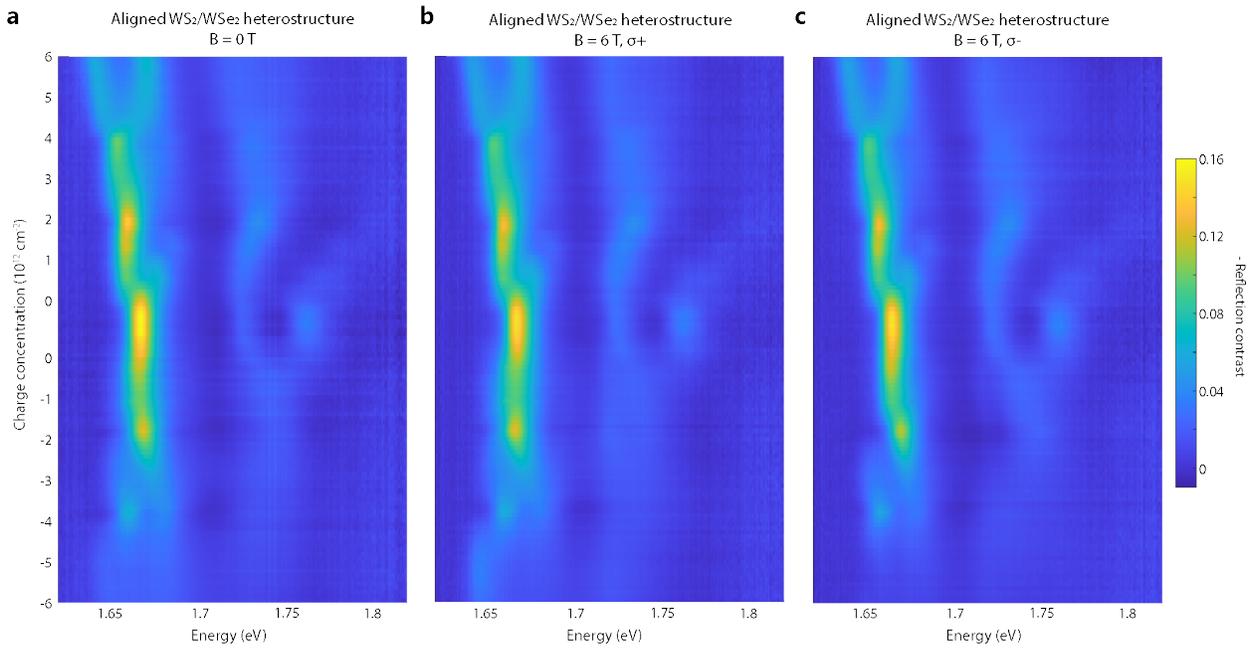

**Fig. S10. | Doping and magnetic field dependent reflection contrast of moiré excitons in aligned WSe$_2$/WS$_2$ heterostructure over extended charge doping range.** (**a**) Doping dependent reflection contrast data at zero magnetic field. (**b**) Doping dependent reflection contrast data at B = 6 T, probing K-valley excitons with σ+ light. (**c**) Doping reflection dependent data at B = 6 T, probing K'-valley excitons with σ- light.

## S11. Linecuts of doping dependent reflection contrast

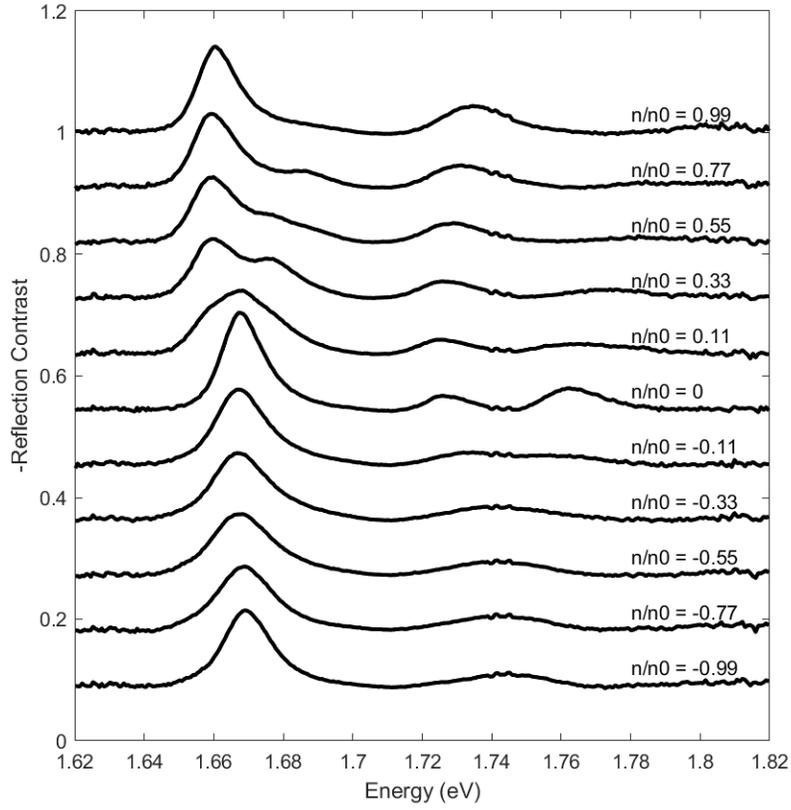

**Fig. S11. | Linecuts of reflection contrast spectra of moiré excitons in aligned WSe$_2$/WS$_2$ heterostructure.**

## S12. Fitting of peak energy in doping dependent reflection contrast

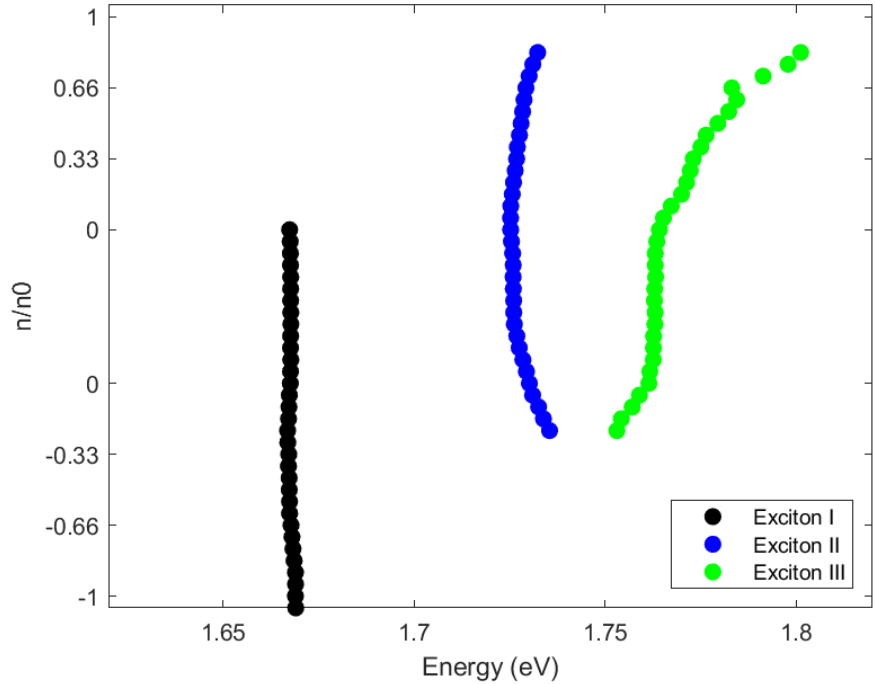

**Fig. S12. | Fitting of peak energies in doping dependent reflection contrast.** At each charge concentration, the spectrum was fit using three Lorentzian functions. The extracted peak energies are plotted only in doping ranges where the peaks can be easily distinguished. As discussed in the main text, peak I is not strongly modified by hole doping, while peak III redshifts with hole doping (scenario B) and blueshifts with electron doping (scenario B). We do not fit peak I under electron doping since it is mixed with peak III (scenario A).

## S13. Doping dependent reflection contrast for additional aligned WSe$_2$/WS$_2$ heterostructure

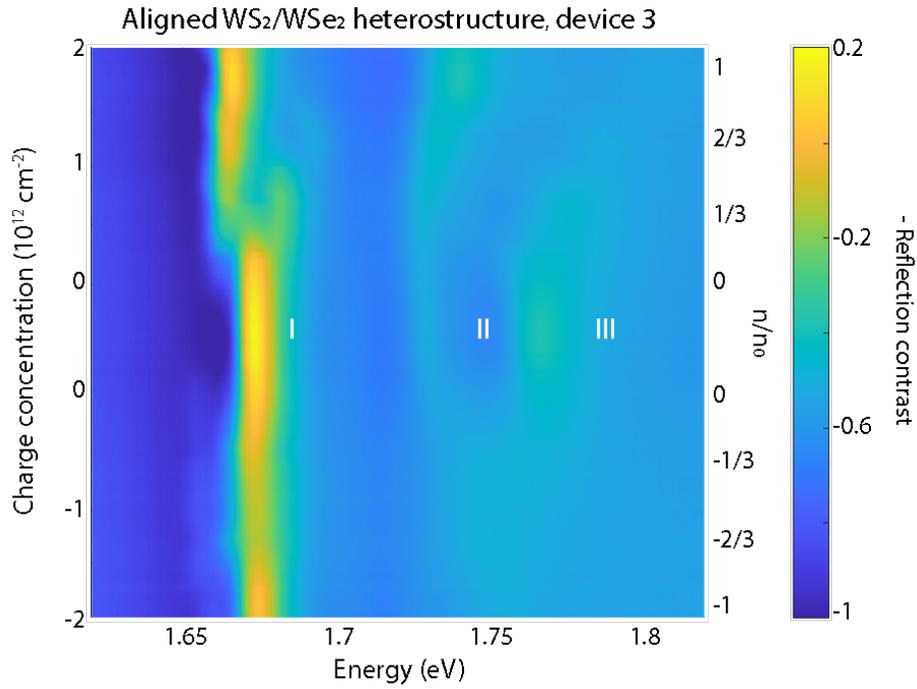

**Fig. S13. | Doping dependent reflection contrast spectra of moiré excitons in a second aligned WSe$_2$/WS$_2$ heterostructure, device 3.** The three moiré exciton states (I, II, III) show very similar doping dependence as the states shown in the main text (device 1).